\documentclass[11pt, oneside]{article}   	
\usepackage{graphicx}
\usepackage{float}
\usepackage[paper=a4paper,left = 20mm, right = 20mm, top = 20mm, bottom = 20mm]{geometry}
\usepackage{authblk}

\usepackage{amssymb, amsmath, amsthm}

\usepackage[utf8]{inputenc}
\usepackage{natbib}
\usepackage{indentfirst}

\usepackage[colorlinks=true,
		    linkcolor=blue,
		    citecolor=blue,
		    urlcolor=blue
		    ]{hyperref}

\def\L{{\cal L}}

\def\pp{\medskip}

\title{\vspace*{-1cm}\bfseries On a method of solving the Black-Scholes Equation}
\author[1]{\normalsize B. Yermukanova}
\author[2]{L. Zhexembay}
\author[3]{N. Karjanto}
\affil[1]{\small{Department of Economics, School of Humanities and Social Sciences\newline Nazarbayev University, Astana,  Kazakhstan}}
\affil[2]{Department of Mathematics, School of Science and Technology, Nazarbayev University, Astana, Kazakhstan}
\affil[3]{Department of Mathematics, University College, Sungkyunkwan University, Natural Science Campus\newline
2066 Seobu-ro, Suwon-si, Jangan-gu, 16419, Gyeonggi-do, Republic of Korea}

\date{}							

\begin{document}
\maketitle
\begin{abstract}
The paper proposes a different method of solving a simplified version of the Black-Scholes equation.
In the first part of the paper, the Black-Scholes equation is transformed into ordinary differential equation to get a solution similar to the solution of the Euler equation. The second part of the paper focuses on partial differential equation. Separation of variables method is used to solve the Black-Scholes equation.
Plots corresponding to put and call options are also given.
\end{abstract}

\section{Introduction}
Differential equations have a great variety of applications in different fields of science such as engineering, physics, biology, pharmacokinetics (\cite{Li:2014}).
Yet, there are only a few of their applications in economics or finance. Particularly, well-known models involving differential equations are only economic growth model and Black-Scholes equation. The latter one will be discussed in the paper. In 1977, Myron Scholes together with Fischer Black were awarded a Nobel Prize in economics for the formulation of stock options formula through ``new method of determining the value of derivative'' (\cite{Jarrow:1999}).
So, Black-Scholes model deals with one of the most important issues in quantitative finance  pricing of options (\cite{Rodrigoa:2006}). This model has significant implications  both theoretical and practical  since finance plays a great role in economies around the world (\cite{Bohner:2009}).

\section{Ordinary Differential Equation}
\subsection{Background information and underlying assumptions}
In practice, Black-Scholes model of option pricing was applied to various ``commodities and payoff structures'' (\cite{dar:2005}).
Black-Scholes model is widely used for American options as well as for European options. Therefore, the model has a wide variety of applications. Before considering Black-Scholes model, there is a number of assumptions that should be made. Fischer Black calls them ``ideal condition'' of the market (\cite{Black:1973}). These assumptions are important to emphasize because it is well-known that stock markets are often volatile compared to other parts of the economy.

There are five underlying assumptions:
\begin{enumerate}
\item First assumption that should be made is information about values of short-term rates is available and short-term interest rates are constant (\cite{Black:1973}).

\item Secondly, stock pays no dividend (\cite{Black:1973}).

\item Thirdly, transaction costs that occur while buying or selling securities are eliminated (\cite{Black:1973}).

\item Fourthly, it is achievable to borrow fraction of price of stock one wants in order to hold the
stock (\cite{Black:1973}).

\item The last assumption states that short selling of a security in necessary situations is allowed in
the market (\cite{Black:1973}).
\end{enumerate}
Thanks to these assumptions, option price will be the function of the time period and stock price
only. In the following paragraphs, option price will be reduced to the function of stock price only for simplicity.

Generally, option values increase when stock prices rise. A positive relationship between option value and stock price may be easily seen from the following graph (\cite{Black:1973}). 
As it can be seen from the figure, graphs representing the relationship between the option price and stock price at different time periods ($T_1, T_2, T_3$) lie below 45-degree line, which shows that option prices are more volatile than the stock prices (\cite{Black:1973}). The volatility of option prices lead to the following statement: if the price of the stock increases by a certain amount, greater percentage change will be generated in option prices.
The graph illustrated below shows what the paper seeks to explain through Black-Scholes model.
\begin{figure}[h]
\centering
\includegraphics[width=0.6\textwidth]{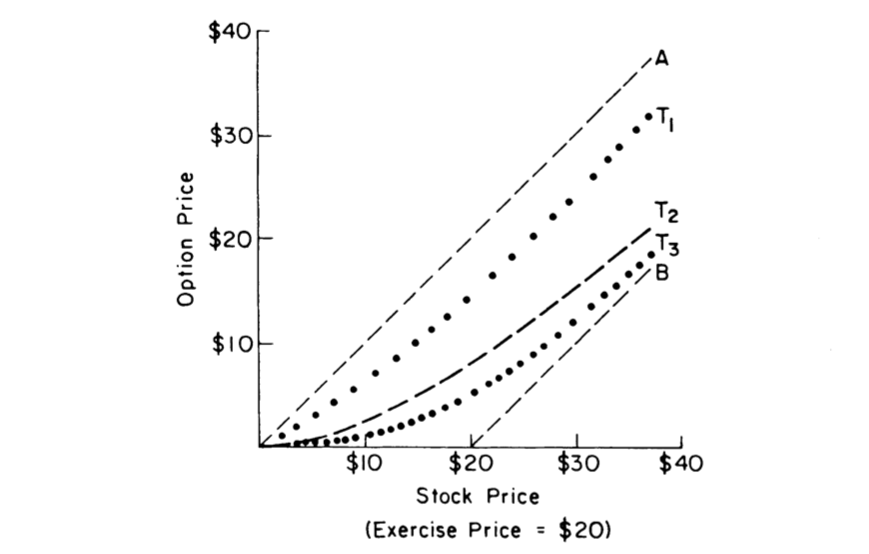}
\caption{The relation between option value and stock price}
\end{figure}

\subsection{Transformation into an Ordinary Differential Equation}
Black-Scholes equation is given by the following expression:   
\begin{equation*}
\frac{\partial C}{\partial t} + \frac{1}{2}\sigma^2s^2\frac{\partial C^2}{\partial s^2} + rs\frac{\partial C}{\partial s} - rC=0,
\end{equation*}
where $C(s, t) =$ price of option, $s =$ price of the stock, $t$ = period of time, $r$ = interest rate (\cite{Company:2007}).
Firstly, it is useful to transform this partial differential equation (PDE) into an ordinary differential equation (ODE) by proposing the following solution: $C(s, t) = C(s)e^{\lambda t}$.
Given that
${\displaystyle \frac{\partial C}{\partial t} = C(s)*\lambda e^{\lambda t}}$
and 
${\displaystyle \frac{\partial C}{\partial s} = \frac{\partial C(s)}{\partial s} e^{\lambda t}}$,
by substituting these equations into the PDE we get:
\begin{equation*}
C(s)*\lambda e^{\lambda t} + \frac{1}{2}\sigma^2s^2\frac{d^2 C(s)}{ds^2}*e^{\lambda t} + rs\frac{dC(s)}{ds}*e^{\lambda t} - rC(s)*e^{\lambda t}=0.
\end{equation*}   
The next step is to rearrange the equation to get second order ODE:
\begin{equation*}
e^{\lambda t} \left[\frac{1}{2}\sigma^2s^2\frac{d^2 C(s)}{ds^2}+ rs\frac{dC(s)}{ds}+C(s)(\lambda - r) \right] = 0.
\end{equation*}       
The latter expression can be reduced to the following equation:     
\begin{equation*}
\frac{1}{2}\sigma^2s^2\frac{d^2 C(s)}{ds^2}+ rs\frac{dC(s)}{ds}+C(s)(\lambda - r)=0
\end{equation*}      
since $e^{\lambda t} \neq 0$.    
     
\subsection{Euler equation}     
To get rid of the coefficient of the first term lets divide everything by 1/2$\sigma^2$:    
\begin{equation*}
s^2\frac{d^2 C(s)}{ds^2}+ \frac{2r}{\sigma^2}*s\frac{dC(s)}{ds}+\frac{2(\lambda-r)}{\sigma^2}C(s)=0.
\end{equation*}   
This equation reminds us the Euler equation:
\begin{equation*}
L(y)=x^2\frac{d^2 y}{dx^2}+\alpha x \frac{dy}{dx}+\beta y = 0.
\end{equation*} 
with real constants $\alpha$ and $\beta$ (\cite{W.:2009}).
 In our case, $\alpha = \frac{2r}{\sigma^2}$ and $\beta = \frac{2(\lambda-r)}{\sigma^2}$, which are positive constants.
Euler equation has the solution of the form
$$y=x^{r_1}+x^{r_2}$$
in case of distinct real roots, and characteristic equation of the form:
\begin{equation*}
F(r) = r(r-1) + \alpha r +\beta =0
\end{equation*} 
(\cite{W.:2009}).

\subsection{Solution of the Black-Scholes equation}
By the assumption given, $\sigma$ and $r$ are positive real numbers because $r$ is an interest rate and $\sigma$ is volatility of the stock as noted earlier in the paper.
Now, a solution in the form of $C(s)=s^k$ can be proposed and applied to the Black-Scholes equation.
The following derivations will be useful in solving our problem:
$$C(s)=s^k, \qquad \qquad \frac{dC(s)}{ds}=k*s^{k-1}, \qquad \qquad \frac{d^2C(s)}{ds^2}=k(k-1)*s^{k-2}.$$
Substituting the derivations back into the earlier equation we get:
\begin{equation*}
\frac{1}{2}\sigma^2s^2k(k-1)*s^{k-2}+ rsk*s^{k-1}+s^k(\lambda - r)=0.
\end{equation*}
The next step is to take $s^k$ out of bracket and derive characteristic equation introduced earlier:
\begin{equation*}
s^k* \left[\frac{1}{2}\sigma^2k(k-1)+ k(r-\frac{1}{2}\sigma^2)+(\lambda - r) \right] = 0.
\end{equation*}

\begin{equation*}
\frac{1}{2}\sigma^2k^2+ rk+(\lambda - r)=0.
\end{equation*}
To find the roots of characteristic equation, let us find discriminant:
\begin{equation*}
D=(r-\frac{1}{2}\sigma^2)^2 - 4* \frac{1}{2}\sigma^2(\lambda - r) = r^2+r\sigma^2 +\frac{\sigma^2}{4} - 2\lambda \sigma^2 >0
\end{equation*}
by assumption.
So, the two distinct roots of characteristic equation will be:
\begin{equation*}
k_{1,2} = \frac{(r-\frac{1}{2}\sigma^2) \pm \sqrt{r^2+r\sigma^2 +\frac{\sigma^2}{4} - 2\lambda \sigma^2}}{\frac{\sigma^2}{2}*2} = \frac{r}{\sigma^2} - \frac{1}{2} \pm
\frac{\sqrt{r^2+r\sigma^2 +\frac{\sigma^2}{4} - 2\lambda \sigma^2}}{\sigma^2}.
\end{equation*}
Therefore, the solution of our problem can be written as:
\begin{equation*}
C(s) = c_1s^{\frac{r}{\sigma^2} - \frac{1}{2} +
\frac{\sqrt{r^2+r\sigma^2 +\frac{\sigma^2}{4} - 2\lambda \sigma^2}}{\sigma^2}} + c_2s^{\frac{r}{\sigma^2} - \frac{1}{2} -
\frac{\sqrt{r^2+r\sigma^2 +\frac{\sigma^2}{4} - 2\lambda \sigma^2}}{\sigma^2}}.
\end{equation*}
The solution represents option value as a function of stock prices. By the assumption $c_1$ and $c_2$ must be positive constants because of a positive relationship between option price and the stock price introduced earlier.
\begin{figure}[H]
\centering
\includegraphics[width=0.6\textwidth]{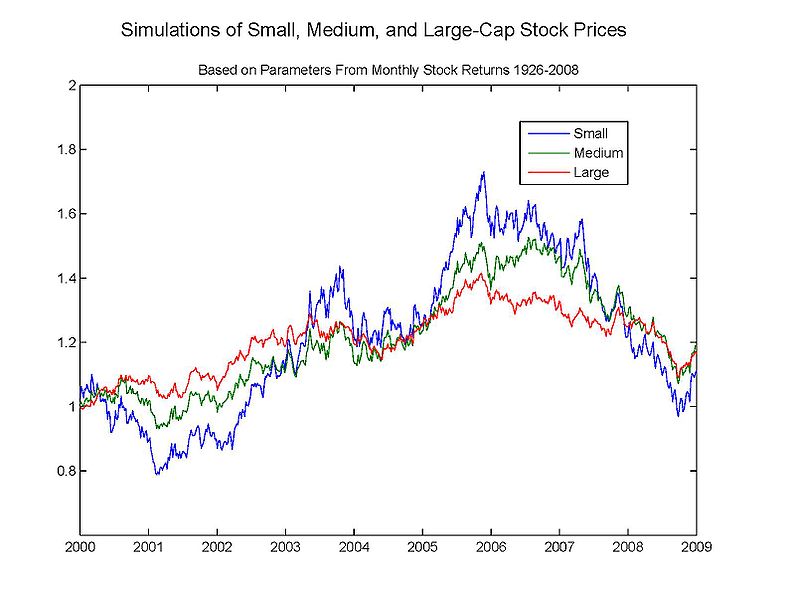}
\caption{Fluctuations in stock prices from 2000 to 2009}
\label{F}
\end{figure}

The figure \ref{F} above shows fluctuations in stock prices from 2000 to 2009 time period. The Black-Scholes presented in the paper is useful to explain, predict and estimate option prices based on stock prices in the financial world. Black-Scholes model gives more accurate estimates of option prices than other earlier developed models because it takes into account such factors influencing the stock prices as transaction costs, riskiness of assets, illiquid markets (\cite{Ankudinova:2008}).
Therefore, the model is used to estimate European call options, which consolidates its role in applied economics (\cite{Barad:2014}).

Black-Scholes model focuses on option or security that is held for a certain period of time and gives the owner right to make market operations such as buying and selling. Two types of securities can be specified: American options and European options (\cite{Black:1973}). The difference of American options from European options is its quality that allows the owner to buy or sell the security until the maturity date, whereas the latter one does not allow conducting market operations until the security matures.

According to empirical tests made by Black and Scholes, estimated option values deviate slightly from what they are in practice. Although those who demand stocks and bonds pay higher (to a small degree) prices for the products in securities markets, suppliers receive payments fairly close to what the formula calculates (\cite{Black:1973}).  This gap occurring between prices paid and received by demanders and suppliers may be understood by transaction costs $-$ costs associated with exchanges $-$ which occur as a result of a variety of services in an industry.

\section{The essence of the Black-Scholes Equation}
There are two types of options that can be specified:``American option'' and ``European option'' (\cite{Black:1973}). American options are the ones that are checkable on demand, particularly they can be returned before the maturity date. Whereas European options are the ones that can be returned only on a specific date, when they mature. So, American options are more liquid than European options.

As noted earlier, an option is more valuable when the stock price is higher (\cite{Black:1973}).  Also, option depends on the maturity date, particularly the date of expiration. If maturity date is over a long period of time, then payments, particularly dividends that are paid on specific periods of time for the option will be less. On the other hand, if maturity date is during a short period of time, then dividends are higher.

Since Black-Scholes equation is a theoretical prediction of stock movements in the market, there are some restrictions that should be noted. Beside theory, there is a real world in which conditions in the stock market may not be that ideal as predicted by the model. Also, the reason theoretical models are built in finance and economics is that it is difficult to test or do an experiment with the real world. For example, in physics the experiments can be done inside the laboratories as well as in chemistry. However, experimenting with financial economics situations imposes high costs as they are closely connected with markets, which exist in our real world. So, making an experiment with the financial markets is difficult because of globality of problems.

The only information for conducting policies or solving problems in finance is past data collection. Data in the past is similar with a tool that helps to understand general patterns of markets, movements of asset prices, behaviors of economic agents, relationships and co-movements of financial variables. 
With data collected in the past, it is possible to draw graphs relating the economic variables, showing their trends and movements. Also, data helps to predict future trend in markets, however only roughly.

Now as we understand the importance of theoretical models, there are some conditions that should be specified regarding the Black-Scholes model. This is a complete list of assumptions additional to that noted earlier in the paper:

\begin{enumerate}
\item  There are perfect information about short - term interest rates and their movements are constant over long period of time (\cite{Black:1973}).
\item  Movement of stock prices is random and its variance is proportional to the square of stock prices; therefore distribution of stock prices is lognormal over some finite period $t$
(\cite{Black:1973}). 
\item Unlike the real world, where stocks pay dividends to the shareholders, in the model
stocks do not pay any payments (\cite{Black:1973}). 

\item Only European options are to be considered in the model, so they are returned on a
maturity date (\cite{Black:1973}). 

\item Unlike in financial markets where transaction costs exist in all operations, it is assumed
that buying or selling stocks do not impose any transaction costs (\cite{Black:1973}). 

\item Any fraction of the price of security can be borrowed at short $-$ term interest rate (\cite{Black:1973}). 

\item There is a possibility of short selling with no penalty or any costs. A seller will accept the
price that buyer tells, agrees to meet at some specific time in the future, and pay the amount equal to that of the security price (\cite{Black:1973}). 
\end{enumerate}

Due to these assumptions, option value will depend only on time and stock price; and other
variables are taken to be constant so as to simplify our model (\cite{Black:1973}). Therefore, the value of the option $w$ reduces to the following simple function:
$$w = f(x, t),$$
where $x$ is the price of the stock and $t$ is the period of time (\cite{Black:1973}). 
The expression above tells us how the value of the option changes if one of the variables, price of the stock or period of time, changes.
The assumption that stocks pay no dividends gives an advantage of getting to more complicated problems with options. One example is under certain conditions, the formula can be applied to American options, which can be issued before maturity date (\cite{Black:1973}). \cite{Black:1973} studied this particular case in 1973.

The Black-Scholes equation helps to calculate not only option value, but also more complicated assets such as warrants, which are liabilities of corporations (\cite{Black:1973}). Warrants are generally considered as options. Also, the value of corporate liabilities may be calculated via the formula. Often corporate liabilities are not viewed as options. We may consider a case of a company, which has assets of shares of another company (\cite{Black:1973}). 
Also, bonds are ``pure discount bonds'', which indicates a bond pays fixed amount of money (\cite{Black:1973}).  Assume maturity of 10 years. Also, assume that a company has a restriction of paying no dividend until the maturity date. Then, assume the company is planning to sell all of its stocks after 10 years and pay the amount for holders of bonds (\cite{Black:1973}). So, these certain conditions let us to calculate a value of corporate liabilities using Black-Scholes equation as the assumptions make corporate liabilities similar to options.

\section{Partial Differential equation}
This part of our paper will greatly consist of the material from the book by \cite{Salsa:2014}.
Also note that \cite{Zhexembay:2016} focus on numerical solution of nonlinear Black-Scholes equation using Finite Element Method.
Let us construct a differential equation that will help us to describe the evolution of $V(s, t).$  The following hypotheses are established:
\begin{itemize}
\item $S$ follows a lognormal law;

\item The volatility $\sigma$ is constant and known;

\item There are no transaction costs or dividends;

\item It is possible to buy or sell any number of the underlying asset;

\item There is an interest rate $r > 0$, for a riskless investment. This means that 1-dollar in a bank at 
time $t = 0$ becomes $e^{rT}$ dollars at time $T$; 

\item The market is arbitrage free.
\end{itemize}

The last hypothesis is crucial in the construction of the model and means that there is no opportunity for instantaneous risk-free profit. 
It could be considered as a sort of conservation law for money!
We can translate this principle into mathematical terms through the notion of hedging and the existence of self-financing portfolios. The fundamental idea is to calculate the return of $V$ through  formula and then to build a riskless portfolio $\Pi$. 
This portfolio contains shares of $S$ and the option. 
$\Pi$ must increase at the current interest rate $r$, i.e. $d\Pi = r\Pi dt$, which turns out to coincide with the fundamental Black-Scholes equation.
Now let us move to the calculation of the differential of $V$ through the means of the formula. Since
\begin{equation*}
dS = \mu Sdt + \sigma dB,
\end{equation*}
the obtained result is
\begin{equation}\label{1}
dV = [V_t+\sigma SV_s + 1/2\mu^2 S^2V_{SS}] dt+\sigma SV_SdB.
\end{equation}

In the formula (\ref{1}) we have risk term $\sigma SV_S dB$. So, our next goal will be an elimination of this term. This can be acquired by establishing a portfolio $\Pi$ consisting of the option and a quantity $-\triangle$ of underlying:
$$\Pi = V - S\triangle.$$
This operation is valuable financial procedure called hedging. Now, let us turn out attention to the particular period of time, say $(t, t+dt)$ during which $\Pi$ goes through a variation $d\Pi$. If we succeed in keeping $\triangle$ equal to its value at $t$ during the interval $(t, t+dt)$, the variation of $\Pi$ is given by
$$d\Pi = dV - \triangle dS.$$
Since the mentioned formula is a cornerstone of the whole construction, it should be explained properly.
Implementing $\ref{1}$ the we acquire:
\begin{equation}\label{2}
d\Pi=dv-\triangle dS = [V_t+\mu SV_s+\frac{1}{2}\sigma^2S^2V_{ss} - \mu S\triangle]dt +\sigma S(V_s - \triangle)dB.
\end{equation}
Thus, if we choose
\begin{equation}\label{3}
\triangle =V_s,
\end{equation}
where  $\triangle$ is the value of $V_s$ at $t$, we eliminate the stochastic component in (\ref{2}). The development of
the portfolio $\Pi$ is now deterministic and its dynamics can be described by the equation:
\begin{equation}\label{4}
d\Pi = [V_t+\frac{1}{2}\sigma^2S^2V_{ss}]dt.
\end{equation}
The choice of (\ref{3}) seems inexplicable, but it can be justified by the fact that $V$ and $S$ are dependent and the random component in their dynamics is proportional to $S$. Thus, we the linear combination of $V$ and $S$ is chosen wisely, such component should vanish.

Now let us implement the no-arbitrage principle. Investing $\Pi$ at the riskless rate $r$, after a time $dt$ we have an increment $r\Pi dt$. Comparison between $r\Pi dt$ and $d\Pi$ is given by (\ref{4}).
If $d\Pi > r\Pi dt$, we borrow an amount $\Pi$ to invest in the portfolio. The return $d\Pi$ would be greater of the cost $r\Pi dt$, so that we make an instantaneous riskless profit
$$d\Pi - r\Pi dt.$$
If $d\Pi < r\Pi dt$, we sell the portfolio $\Pi$ investing it in a bank at the rate $r$. This time we would make an instantaneous risk free profit
$$r\Pi dt - d\Pi.$$
Therefore, the arbitrage free hypothesis forces
\begin{equation}\label{5}
d\Pi = [V_t+\frac{1}{2}\sigma^2S^2V_{ss}]dt = r\Pi dt .
\end{equation}
Substituting $\Pi = V - S\triangle = V-V_sS$
into (\ref{5}), we obtain famous Black-Scholes equation:
\begin{equation}\label{6}
\L V = V_t+\frac{1}{2}\sigma^2S^2V_{ss}+rSV_s -rV=0.
\end{equation}
 Since the coefficient of $V_{ss}$ is positive, (\ref{6}) is a backward equation. In order to get well-posed problem, we need to impose final condition (at $t = T$), a side condition at   $
S=0$ and one condition for $S \rightarrow +\infty$.

\begin{itemize}
\item{Final conditions $(t=T)$}

Call. If at time $T$ we have $S>E$ then we exercise the option, with a profit $S-E$.
If $S \le E$, we do not exercise the option with no profit. The \textit{final payoff} of the option is therefore
$$C(S, T) = \text{max}[S-E, 0] = (S-E)^+, S>0.$$

Put. If at time $T$ we have $S \le E$, we do not exercise the option, while we exercise the option if $S < E$. The \textit{final payoff} of the option is therefore
$$P(S,T) = \text{max}[E-S, 0] = (E-S)^+, S>0.$$

\item{Boundary conditions ($S=0$ and  $S \rightarrow +\infty$)}

Call. If $S=0$ at a time $t$, $S=0$ thereafter, and the option has no value; thus 
$$C(0, t) = 0, t\ge 0.$$
As $S \rightarrow +\infty$, at time $t$, the option will be exercised and its value becomes practically equal to $S$ minus the discounted exercise price, so
$$C(S,t) - (S-e^{-r(T-t)}E) \rightarrow 0 \ \text{as} \ S \rightarrow +\infty.$$

Put. If at a certain time is $S=0$, so that $S=o$ thereafter, the final profit is $E$. Thus, to
determine $P(0,t)$ we need to determine the present value of $E$ at time $T$, that is
$$P(S, t) = Ee^{-r(T-t)}.$$

If $S \rightarrow +\infty$, we do not exercise the option, hence
$$P(S, t) = 0 \ \text{as} \ S \rightarrow +\infty.$$
\end{itemize}

\textit{Solution of the Black-Scholes equation}\pp

Let us summarize our model in the two cases.\pp
\begin{itemize}
\item{Black-Scholes equation}
\begin{equation}\label{BS}
V_t+\frac{1}{2}\sigma^2S^2V_{ss}+rSV_s -rV=0.
\end{equation}

\item{Final payoffs}
$$C(S, T) = (S-E)^+ \qquad \text{(call)} $$
$$P(S,T) =(E-S)^+   \qquad \text{(put)}. $$

\item{Boundary conditions}
$$C(S,t) - (S-e^{-r(T-t)}E) \rightarrow 0 \qquad \qquad \text{as} \qquad S \rightarrow +\infty \qquad \text{(call)}$$
$$P(0, t) = Ee^{-r(T-t)}, \qquad P(S, t) = 0     \qquad \text{as} \qquad S \rightarrow +\infty \qquad \text{(put)}.$$
\end{itemize}

The problem above can be simplified to a global Cauchy problem for the heat equation. Thus, the explicit formulas for the solutions can be obtained. Firstly, a change of variables should be performed so as to reduce the Black-Scholes equation to constant coefficients and to pass from backward to forward in time.  
Note also that $1/ \sigma^2$ can be considered an intrinsic reference time while the exercise price $E$ gives a characteristic order of magnitude for $S$ and $V$. Thus, $1/ \sigma^2$ and $E$ can be used as rescaling factors to introduce a dimensionless variable.
Let us set
$$x= \ln \left(\frac{S}{E} \right), \qquad 
\tau = \frac{1}{2} \sigma^2 (T-t),  \qquad  \omega(x, \tau) = \frac{1}{E}V(Ee^x, T-\frac{2\tau}{\sigma^2}).$$
When $S$ goes from $0$ to $=\infty$, $x$ varies from $-\infty$ to $+\infty$. 
When $t=T$ we have $\tau = 0$. 
Moreover: 
$$V_t = -\frac{1}{2}\sigma^2E\omega_\tau$$
$$V_s = \frac{E}{S}\omega_x, V_{ss} = -\frac{E}{S^2}\omega_x+\frac{E}{S^2}\omega_{xx}.$$
If we put this into (\ref{BS}) we will end up with the following:
$$-\frac{1}{2}\sigma^2\omega_\tau+\frac{1}{2}(-\omega_x+\omega_{xx}) +r\omega_x - r\omega = 0$$                   
or
$$\omega_\tau=\omega_{xx} +(k-1)\omega_x - k\omega$$
where $k=\frac{2r}{\sigma^2}$ is a parameter with no dimension. If we set
$$\omega(x, \tau) = e^{-\frac{k-1}{2}x - \frac{(k+1)^2}{4}\tau}\nu(x, \tau),$$
we find that $\nu$ satisfies
$$\nu_\tau - \nu_{xx} = 0, \ -\infty<x< +\infty, \ 0 \le \tau \le T.$$

It is worth mentioning that the final condition for $V$ is an initial condition for $\nu$. Performing
following several steps we find out that
$$\nu(x, 0) = g(x) =
\left\{
\begin{array}{l}
\displaystyle{e^{\frac{1}{2}(k+1)x} - e^{\frac{1}{2}(k-1)x},} \  x>0\\
0, \ x \le 0 
\end{array}
\right.$$ 
for the call option, and
$$\nu(x, 0) = g(x) =
\left\{
\begin{array}{l}
\displaystyle{e^{\frac{1}{2}(k-1)x} - e^{\frac{1}{2}(k+1)x},} \  x<0\\
0, \ x \ge 0 
\end{array}
\right.$$ 
for the put option.

Now we can use the preceding results to derive the formula of the solution. The solution is unique and it is given by formula
$$\nu(x, \tau) = \frac{1}{\sqrt{4\pi \tau}} \int_R^\infty g(y)e^{-\frac{(x-y)^2}{4\tau}}dy.$$
To have a more general formula, let $y=\sqrt{2\tau z}+x$. Then, focusing on the call option:
\begin{eqnarray*}
\nu(x, \tau) &=& \frac{1}{\sqrt{4\pi \tau}} \int_R^\infty g(\sqrt{2\tau z}+x)e^{\frac{-x^2}{2}}dy \\
&=& \frac{1}{\sqrt{2\pi}} \left[ \int_{-\frac{x}{\sqrt{2\tau}}}^\infty 
e^{\frac{1}{2}(k+1)(\sqrt{2\tau z}+x) - \frac{1}{2}z^2}dz - \int_{-\frac{x}{\sqrt{2\tau}}}^\infty 
e^{\frac{1}{2}(k-1)(\sqrt{2\tau z}+x) -\frac{1}{2}z^2}dz  \right].
\end{eqnarray*}
After modifying those two integrals, we get
$$\nu(x, t) = e^{\frac{1}{2}(k+1)x+\frac{1}{4}(k+1)^2 \tau} N(d_+) -e^{\frac{1}{2}(k-1)x+\frac{1}{4}(k-1)^2 \tau} N(d_-) $$
where
$$N(z)=\frac{1}{\sqrt{2\pi}}\int_{-\infty}^z e^{-\frac{1}{2}y^2} dy$$
is the distribution of a standard normal random variable and
$$d_\pm=\frac{x}{\sqrt{2\tau}}+\frac{1}{2}(k\pm 1)\sqrt{2\tau}.$$
Returning to the original variables, for the call we have:
$$C(S, t) = SN(d_+) - Ee^{-r(T-t)}N(d_-)$$
with 
$$d_\pm=\frac{\ln(S/E) +(r \pm \frac{1}{2}\sigma^2)(T-\tau)}{\sigma \sqrt{T-\tau}}.$$
The formula for the put is
$$P(S, t) = Ee^{-r(T-t)}N(d_-) - SN(d_+).$$

It can be shown that
$$\triangle = C_s = N(d_+) > 0  \quad \qquad \text{for the call}  $$
$$\triangle = P_s = N(d_+) - 1 < 0    \qquad \text{for the put} \ $$ (\rm \cite{Salsa:2014}).
We should pay particular attention to the fact that both $C_s$ and $P_s$ are strictly increasing with respect to $S$. Thus, the functions $C, P$ are strictly convex functions of $S$, for every $t$, namely $C_{ss} > 0$ and $P_{ss} > 0$.

\begin{itemize}
\item Put-call parity. Put and call options with the same exercise price and expiry time can be connected by forming the following portfolio:
\end{itemize}
$$\Pi = S+P-C$$
where the minus in front of $C$ shows \textit{short position} (negative holding).
For this portfolio the final payoff is
$$\Pi (S, T) = S+(E-S)^+ - (S-E)^+.$$

\begin{figure}[H]
\centering
\includegraphics[width=0.6\textwidth]{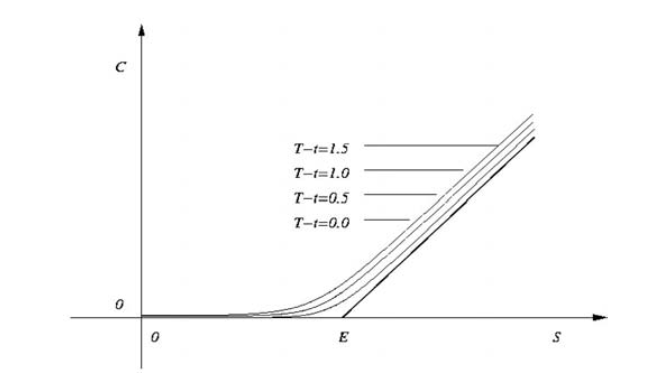}
\caption{The value for the European call option}
\label{EC}
\end{figure}

\begin{figure}[H]
\centering
\includegraphics[width=0.6\textwidth]{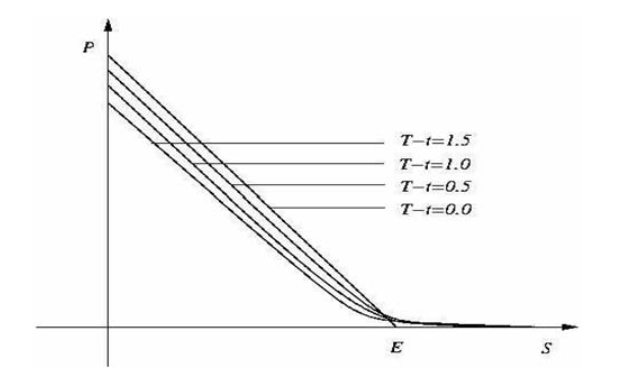}
\caption{The value function of the European put option}
\label{EP}
\end{figure}

If $E \ge S$, we have 
$$\Pi (S, T) = S+(E-S) - 0 = E,$$
while if $E \le S$
$$\Pi (S, T) = S+ 0 - (S-E) = E.$$
Therefore, at expiry, the payoff is always equal to $E$ and it forms a riskless profit, whose value at $t$ must be equal to the discounted value of $E$, since the no-arbitrage condition was
imposed. So, we find the subsequent relation (\textit{put-call parity})
\begin{equation}\label{PC}
S+P-C = Ee^{-r(T-t)}.
\end{equation}
Formula (\ref{PC}) reveals that, with the value of $C$ (or $P$) available, the value of $P$ (or $C$) can be obtained.
From (\ref{PC}), since $Ee^{-r(T-t)} \le E$ and $P \ge 0$, we get
$$C(S, t) = S+P - Ee^{-r(T-t)} \ge S-E$$
and since $C \ge 0,$
$$C(S, t)  \ge (S-E)^+.$$
It can be observed that the value of $C$ is always greater than the final payoff. However, this
property does not hold for a put. In fact,
$$P(0, t) =  Ee^{-r(T-t)} \le E$$
so the value of $P$ is less than the final payoff when $S$ approaches $0$, and it is greater just before expiring. The figures \ref{EC} and \ref{EP} demonstrate that.

\begin{itemize}
\item Different volatilities. The comparison between the value of two options with the different volatilities $\sigma_1$ and $\sigma_2$ can be conducted through the means of the maximum principle. Let us assume that the exercise price and strike time are the same for the both cases, $E$ being the exercise price and $T -$ strike time. Another assumption that we allow is that  $\sigma_1 > \sigma_2$. Denote the values of the respective call options $C^{(1)}, C^{(2)}$. With decreasing amount of risk the value of the option should decline as well. What we want to show is that
\end{itemize}
$$C^{(1)} > C^{(2)}, \quad S>0, \quad 0 \le t \le T.$$
Let $W = C^{(1)} - C^{(2)}$ . Then
\begin{equation}\label{W}
W_t +\frac{1}{2}\sigma^2S^2W_{ss}+rSW_s -rW=\frac{1}{2}(\sigma_2^2 -  \sigma_1^2)S^2C_{ss}^{(1)}.
\end{equation}
with $W(S, T) = 0, \ W(0,t ) = 0$ and $W \rightarrow 0$ as $S \rightarrow +\infty.$
It is obvious that (\ref{W}) is a nonhomogeneous equation, with the right hand side being negative
for $S>0$, since $ C_{ss}^{(1)} > 0$. We know that $W$ is continuous in the half strip   $[0,+\infty) \times [0, T]$ and
disappears at infinity, it reaches its global minimum at  $(S_0, t_0)$. 

We claim that the minimum is zero and cannot be obtained at a point in the intervals $(0,+\infty) \times [0, T)$.
The equation is backward, so $t_0= 0$ is not considered. Assume that $W(S_0, t_0) \le 0$ with $S_0 > 0$ and $0<t_0<T.$ Thus,
$$W_t(S_0, t_0) = 0$$
and 
$$W_s(S_0, t_0) = 0, \qquad  W_{ss}(S_0, t_0) \ge 0.$$
Substituting $S=S_0, \ t=t_0$ into (\ref{W}) we observe a contradiction. Thus, $W = C^{(1)} - C^{(2)} > 0$ for $S>0, \ 0<t<T$.

In 1972, empirical tests on call-options by Fischer Black and Myron Scholes were done (\cite{Black:1973}).The results of the tests show that actual prices at which agents of the economy buy and sell options deviate systematically from the prices predicted by the Black- Scholes model 
(\cite{Black:1973}). Options are bought at consecutively higher prices from those predicted by the model, whereas options are sold approximately at the prices predicted by the model (\cite{Black:1973}). It should be noted that the difference in prices paid by option buyers are higher for lower-risk stocks than for higher-risk stocks (\cite{Black:1973}). The latter point makes sense because low-risk stocks are always preferable than high-risk stocks since the probability that low-risk stocks generate large profits and do not default are higher. For option buyers, there are high transaction costs involved in the real world. 
This fact might explain why formula underestimates the price paid by option buyers because the model is built under assumption of no transaction costs. 
According to \cite{Black:1973}, getting into account the magnitude of the transaction costs in the market, misestimation of the prices does not indicate potential profit opportunities for speculators in the market.

\section{Separation of variables method}
This section considers the partial Black-Scholes equation of the form 
$$\frac{\partial C}{\partial t}+\frac{1}{2}\sigma^2s^2\frac{\partial^2 C}{\partial s^2} +rs\frac{\partial C}{\partial s} = 0.$$
The aim of the section is to introduce separation of variables method in order to solve the equation and find the general solution.
Let the function $C(s, t)$ to be written in the following form:
$$C(s, t)=S(s)T(t).$$
Then, the following partial derivatives can be derived:
$$\frac{\partial C}{\partial t} = ST',$$
$$\frac{\partial C}{\partial s} = TS',$$
$$\frac{\partial^2 C}{\partial t^2} = S''T.$$
Substituting the derivatives back to the original equation, we get:
$$ST'+\frac{1}{2}\sigma^2s^2S''T+rsS'T-rST= 0.$$
For simplicity, let $\frac{1}{2}\sigma^2 = a$ and $r=b$ for now.
Dividing the equation by $ST,$
$$as^2\frac{S''}{S} +bs\frac{S'}{S}+\frac{T'}{T} - b= 0.$$
Rearranging and equating to a constant ($c>0$), we get pair of two ordinary differential equations:
$$as^2\frac{S''}{S} +bs\frac{S'}{S}=b-\frac{T'}{T} = c.$$

The first equation
$$as^2S'' +bsS' = cS$$
can be solved by Euler equation method introduced earlier.
Dividing both sides of the equation by $a$ and rearranging:
$$s^2S'' +\frac{b}{a}sS' - \frac{c}{a}S = 0.$$
Characteristic equation is therefore:
$$d(d-1) +\frac{b}{a}d - \frac{c}{a} = 0$$
or
$$d^2 +(\frac{b}{a} -1)d - \frac{c}{a} = 0.$$
The characteristic equation has the following solutions (assuming $D > 0$):
$$d_{1,2} = \frac{(1-\frac{b}{a}) \pm \sqrt{(\frac{b}{a} -1)^2 - 4*1*(-\frac{c}{a})}}{2}.$$
Therefore, the general solution of the equation is
$$S(s) = s^{\frac{(1-\frac{b}{a}) + \sqrt{(\frac{b}{a} -1)^2 - 4(-\frac{c}{a})}}{2}}
+s^{\frac{(1-\frac{b}{a}) - \sqrt{(\frac{b}{a} -1)^2 - 4(-\frac{c}{a})}}{2}}.$$
Substituting respective expressions instead of $a$ and $b$, we get the solution in explicit form:
$$S(s) = s^{\frac{(1-\frac{2r}{\sigma^2}) + \sqrt{(\frac{2r}{\sigma^2}-1)^2 - 4(-\frac{2c}{\sigma^2})}}{2}}
+s^{\frac{(1-\frac{2r}{\sigma^2}) - \sqrt{(\frac{2r}{\sigma^2}-1)^2 - 4(-\frac{2c}{\sigma^2})}}{2}}.$$

Next, the aim is to find the solution of $T(t)$ from: 
$$b-\frac{T'}{T} = c.$$
$$T' - (b-c)T = 0.$$
So, the solution of the equation is therefore:
$$T(t) = e^{(b-c)t}.$$
Since the two solutions of the form $S(s)$ and $T(s)$ are found, the final solution of the Black-Scholes equation
$$\frac{\partial C}{\partial t}+\frac{1}{2}\sigma^2s^2\frac{\partial^2 C}{\partial s^2} +rs\frac{\partial C}{\partial s} = 0$$
can be expressed as
$$C(s, t)=S(s)T(t) = \Bigl[s^{\frac{(1-\frac{2r}{\sigma^2}) + \sqrt{(\frac{2r}{\sigma^2}-1)^2 + 2\frac{c}{\sigma^2}}}{2}}
+s^{\frac{(1-\frac{2r}{\sigma^2}) - \sqrt{(\frac{2r}{\sigma^2}-1)^2 +2\frac{c}{\sigma^2}}}{2}}\Bigr] e^{(b-c)t}.$$

\begin{figure}
\centering
\includegraphics[width=0.6\textwidth]{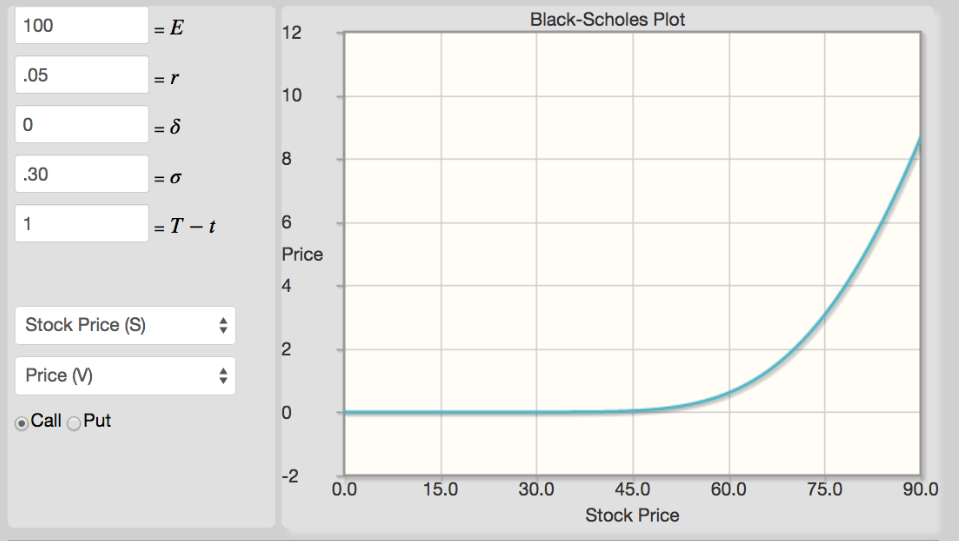}
\caption{Call option}
\label{CO}
\end{figure}

\begin{figure}
\centering
\includegraphics[width=0.6\textwidth]{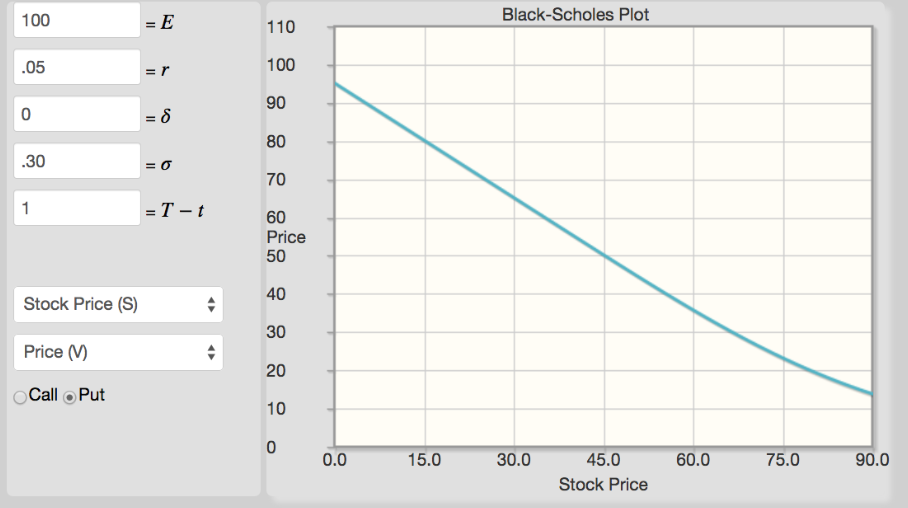}
\caption{Put option}
\label{PO}
\end{figure}

\begin{itemize}
\item{Constructing a plot}
\end{itemize}
Nowadays, there are lots of opportunities to see the graphical solution of the Black-Scholes equation. We decided to rely on one of them and show the plot of this equation.
Statistic Online Computational Research (SOCR) allowed us to construct these two graphs.
In the graph \ref{CO} the Exercise Price $E$ was taken to be equal to 100, Interest Rate   $r= 0.5$, Dividend Rate   $\delta  = 0$, Variance   $\sigma  = 0.3$ and Time to Expiry   $T-t= 1$. The graph was plotted with respect to the Stock Price $S$ and Price $V$.
On the graph \ref{PO} the values of the variables were assumed to be the same and the plot was made according to the Put option.

\section{Conclusion}
To conclude, Black-Scholes model is highly appreciated in quantitative finance because of its accurate and useful estimation of stock prices. Black-Scholes equation represents derivation of option pricing though taking into account such factors as time period $t$, risk-free interest rate $r$ and volatility of stock prices $\sigma$ (\cite{Sheraza:2014}).
Derived solution for the option value is closely related to corporate liabilities, therefore, the formula derived may be used to securities, including common stock and bond (\cite{Black:1973}).
This feature of Black-Scholes model illustrates its flexibility and efficiency of being applied to different contexts in the financial world.
In this paper, we proposed a new method of solving the famous Black-Scholes Equation. Separation of variables method was used to derive a solution to the partial differential equation.

\small{
\bibliographystyle{elsarticle-harv}
\bibliography{arxiv}
}

\end{document}